# Hippocampus segmentation in magnetic resonance images of Alzheimer's patients using Deep machine learning

Hossein Yousefi-Banaem[1], Saber Malekzadeh[2]

[1] Skull Base Research Center, Loghman Hakim Hospital, Shahid Beheshti University of Medical Sciences, Tehran, Iran

[2] School of Data, Khazar University, Baku, Azerbaijan

ORCID number of Hossein Yousefi-Banaem: 0000-0002-3754-4496

ORCID number of Saber Malekzadeh: 0000-0003-3503-8992

**Abstract**

Background: Alzheimer's disease is a progressive neurodegenerative disorder and the main cause of dementia in aging. Hippocampus is prone to changes in the early stages of Alzheimer's disease. Detection and observation of the hippocampus changes using magnetic resonance imaging (MRI) before the onset of Alzheimer's disease leads to the faster preventive and therapeutic measures.

Objective: The aim of this study was the segmentation of the hippocampus in magnetic resonance (MR) images of Alzheimer's patients using deep machine learning method.

Methods: U-Net architecture of convolutional neural network was proposed to segment the hippocampus in the real MRI data. The MR images of the 100 and 35 patients available in Alzheimer's disease Neuroimaging Initiative (ADNI) dataset, was used for the train and test of the model, respectively. The performance of the proposed method was compared with manual segmentation by measuring the similarity metrics.

Results: The desired segmentation achieved after 10 iterations. A Dice similarity coefficient (DSC) = 92.3%, sensitivity = 96.5%, positive predicted value (PPV) = 90.4%, and Intersection over Union (IoU) value for the train (0.9294) and test (0.9293) sets were obtained which are acceptable.

Conclusion: The proposed approach is promising and can be extended in the prognosis of Alzheimer's disease by the prediction of the hippocampus volume changes in the early stage of the disease.

**Introduction**

Alzheimer's disease (AD) as a progressive neurodegenerative disorder leads to dementia in aging by progressive loss of memory and deterioration of cognitive functions [1, 2]. Today, the quality of life of many people is directly or indirectly impaired with the increasing trend of aging in society and the involvement of people with AD and mental retardation. The hippocampus is the gray matter of the temporal lobe on each side of the brain that involves during AD onset [3, 4]. Detection of the hippocampus changes is essential in the early stages of AD onset [3]. The hippocampus has several structures that connected together and other parts of the brain by entorhinal cortex, causing the complex and non-uniform structures [5].

MRI is a method for evaluation of the hippocampus shape and volume [6, 7]. The research on the MR images has shown that the volume of the hippocampus is significantly reduced in the AD [8]. The hippocampus volume can be examined quantitatively or qualitatively [9]. Qualitative examination is performed by a specialist and depends on his/her experience. This type of examination, which is done visually, is prone to the errors [6, 9]. Therefore, in order to quantify the hippocampus volume, different segmentation methods have been proposed [9].

The manual segmentation is a simple method used as a gold standard for quantifying of the hippocampus volume. However, it is a time consuming, subjective and tedious task [10, 11]. Automatic methods can reduce personal dependency of segmentation and increase accuracy. Since, the hippocampus in the MR images is very similar to that of the adjacent structures and there is no definite boundary between them, automatic segmentation is also problematic. On the other hand, the partial volume effect and signal inhomogeneity of the MR images are also problematic [9].

Atlas-based methods are two types; single atlas or multiple atlases. Although single atlas is a simple and fast method [9], the main drawback with this method is the lack of the accurate segmentation results in the variable anatomies. [12]. In contrast, the multi-atlas method can segment the subjects with the different anatomy but it is complex and requires more computation costs. Deformable models are another methods, which are mostly used for medical images segmentation. These methods have less complexity but requires an initial contour to segment the object. Therefore, the results depend on the initial contour location. Another disadvantage of deformable models is degradation of the results by noise [9].

Machine learning refers to developing computer programs that can access data and use it to learn. Machine learning as an application of artificial intelligence (AI) allows systems to automatically learn and progress from experience without explicit planning. In medical image analysis, machine learnng algorithms provide an improved determination of boundary recognition. There are different methods of machine learning [13]. Deep leaning (DL) is one of these methods [14]. Convolutional neural networks (CNN) is the most well-known DL algorithm due to the wide range of the applications in recognizing the different patterns. To date, CNN is the most successful model for use in field of the analyzing medical images. CNN is made up of several layers, in which input information such as an image, produce an output such as the presence or absence of a disease. In the field of the medical image analyses, DL is used to classification, detection, registration and segmentation of an organ or a lesion and the related structure [13].

CNN has been used for the gradation [15] and segmentation [16] of the brain tumors, diagnosis of the cerebral micro-bleeding in the MR images [17] and classification of the pulmonary nodules on the chest radiography [18]. Zhao et al. proposed a multi-scale CNNs model that showed the advantages and superiority compared to the traditional CNNs and other methods. According to their study, the multi-scale CNNs model required more use of the self-learning property, extracting proper boundary features and discriminating fuzzy points to improve the segmentation accuracy [16]. Segmentation of the hippocampus in the MR images of the infant brain using Stacked Auto Encoder (SAE) based on the unsupervised deep learning method was performed by Guo et al. The results showed a Dice ratio of 70.2% for the segmentation of the infant hippocampi. The study required to improve the accuracy and longitudinal consistence because of using an unsupervised DL method [19]. An end-to-end trainable automatic hippocampus segmentation network was proposed by Chen et al. which was based on the hardwiring of CNN and recurrent neural networks (RNN) models. They combined the modified U-Net (U-Seg-Net) and convolutional long short-term memory (CLSTM) to restore the 3D contextual features back to the 2D CNN model. Their results showed that the proposed model indeed achieved with a very high accuracy and consistency without a significant increase in the complexity and computation cost [20]. All of these studies showed that CNNs have a higher accuracy and precision in classification, detection and segmentation of the hippocampus compared to the other methods.

So far, the different tools have been used for the hippocampal segmentation such as FSL FIRST and FreeSurfer. However, these tools are relatively slow. For instance, FSL FIRST takes more than 10 minutes to execute, and FreeSurfer takes more than 4 hours to calculate the hippocampus volume [21]. To reduce the computation cost and improve the segmentation results accuracy, in the current study, we suggested to apply supervised deep learning method to segment the hippocampus in the series of the brain MR images.

The purpose of this study was to understand the concept of the DL, architectural design with using CNNs to obtain the volume of the hippocampus, and segment the structure of the hippocampus in patients before beginning Alzheimer's diseases with an acceptable accuracy, precision and speed.

**Materials and methods**

U-Net architecture of convolutional neural network was used to segment the hippocampus in the brain MRI. For this purpose, following steps were done:

- The data was collected and preprocessed.

- The model was trained by the manual segmented masks.

- This model was tested by the subjects that were not involved in the training process.

The block diagram of the method is shown in figure 1.

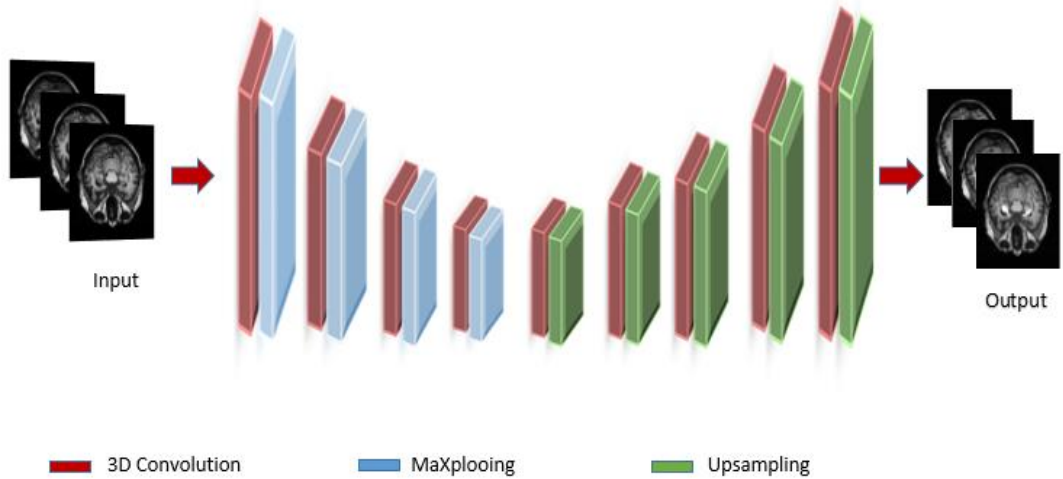

**Figure 1:** General architecture of the proposed method

**Data**

The current study was designed based on the work of Boccardi et al. [22] which was published on their website (http://www.hippocampal-protocol.net/SOPs/index.php). The work presented manual segmentation of MR images of the 135 patients which were available in the Alzheimer's disease Neuroimaging Initiative (ADNI) database (adni.loni.usc.edu). The manual hippocampal segmentation of the images had been done by using Hippocampal Harmonized Protocols (HHP), which is the accepted standard for the hippocampal segmentation [22]. All of the images had been acquired using a T1w_MP-RAGE protocol with a slice thickness of 1.2 mm in a sagittal orientation [22]. The data in ADNI were classified to the four groups based on the clinical status including the AD patients, mild cognitive impairment (MCI) converters, MCI non-converters and healthy controls [23]. The N = 100 of this data was used for the train the model and N = 35 was used for the test of the model. More information including the scanner manufacturer, field magnitude, diagnosis and age bins are seen in Table 1.

**Table 1:** Data description (scanner, field magnitude, group and age).

| Scanner | Siemens<br>N = 23 | GE<br>N = 24 | Philips<br>N = 21 | Siemens<br>N = 23 | GE<br>N = 22 | Philips<br>N = 22 |
|---|---|---|---|---|---|---|
| Field magnitude | 1.5 T | 1.5 T | 1.5T | 3 T | 3 T | 3 T |
| Group* | CTRL<br>N = 22 | MCI<br>N = 24 | AD<br>N = 22 | CTRL<br>N = 22 | MCI<br>N = 22 | AD<br>N = 23 |
| Age** | 60-70<br>N = 19 | 70-80<br>N = 30 | 80 +<br>N = 19 | 60-70<br>N = 21 | 70-80<br>N = 25 | 80 +<br>N = 21 |

\* CTRL: controls, MCI: mild cognitive impairment, AD: Alzheimer's disease

\*\* 80 +: The ages above 80 years old

**Preprocessing**

In the first phase of the study, the MR images of brain with the hippocampal mask and Medical Imaging NetCDF (MINC) format were downloaded and stored. The dimension of each patient's data was 256×256×128 (figure 2A). To reduce the complexity and computational cost, the slices from the image series were selected that were included the hippocampus structure. However, to ensure full coverage of the interested area and reduce the error, several neighboring slices were also selected. Therefore, the data dimension of 128×128×64 (figure 2B) was obtained for the image series. Since the right and left hippocampus regions had been separated in the mask images, the images of each side were read and both side masks were added together to have the complete structure of the hippocampus in one image (figure 2C).

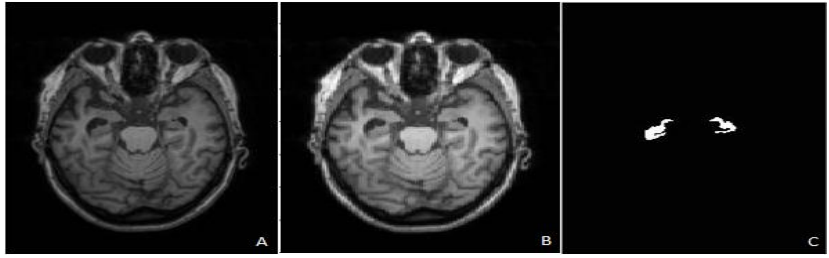

**Figure 2:** Input data. A: Original image with a resolution of 256×256, B: Reduced image size with a resolution of 128×128, C: The left and right mask images added together.

**Training and data**

We used the supervised algorithm for machine learning. The supervised learning system is presented with a dataset $D = \{x, y\}_{n=1}^{N}$ of input features x and label pairs y [13]. The training sets had a pair of the images consist of the hippocampal region and corresponding mask images. The model was learned by analyzing a large number of the training data included the 6336 images of the 100 patients.

The training procedure was implemented on Python 3.7.0, the tensor flow platform and Google Colab environment. This environment eliminated the necessity to use a high graphic processing unit (GPU) computer due to using a GPU NVidia k80 in default. All the hippocampal training and testing procedures were done online in this environment.

We used 10 layers to train our network that each layer consisted of several 3D convolution filters. In the first layer, ten 3D convolution filters with a size of 3×3×3 and s padding = same were applied. The activation function Leaky ReLU with an alpha of 0.3 and a batch size of 1 were used. The batch normalization was performed to prevent over-fitting and increase the accuracy. Max pooling layer with a size of 2×2×2 was applied to reduce the size of the output image by 50%. In each layer, the number of the convolution filters was increased compared to the previous layer, so that, in the sixth layer, the number of the filters reached to 512. From the seventh layer, the number of the filters was decreased and in the tenth layer as fully connected layer, a one-dimensional filter was applied. In the last layer, the sigmoid activation function was used. In addition, Adam optimization function with a learning rate (lr) = $1 \times 10^{-2}$ and a batch size of 1 were applied.

To evaluate the performance of the proposed method, the Dice similarity coefficient (DSC), sensitivity, positive predicted value (PPV) and Intersection over Union (IoU) were calculated as follow:

$$DSC = \frac{2TP}{2TP+FP+FN}, \text{Sensitivity} = \frac{TP}{TP+FN}, PPV = \frac{TP}{TP+FP} \quad [24]$$

$$\text{IoU} = \left|\frac{TP}{FP+TP+FN}\right| \quad [25]$$

TP: The true positives, FP: The false positives, FN: The false negatives [24, 25].

Validation of the proposed method was carried out using IoU metric. For an object in the image, the similarity between the region of interest (ROI) and the ground-truth is determined by IoU which is between 0 and 1 [25].

**Results**

The hippocampus was segmented in the real 3D MR image series using a CNN algorithm depicted in Figure 1. The segmented images were compared with the manual segmentation as a gold standard to validate the method.

The designed convolutional neural network was trained with 10 iterations. In the first iteration, the IoU value for the train and test sets was 0.8222 and 0.9455, respectively. As the learning process progressed, the results were improved which indicates a correct network training. In the last iteration, the IoU value for the train and test sets reached to 0.9294 and 0.9293, respectively. The evaluated results using four metrics of DSC, sensitivity, PPV and IoU are shown in table 2.

Table 2. Evaluation of the proposed method results using four metrics including DSC, Sensitivity, PPV and IoU.

|   | Metric      | Performance (%) |
|---|-------------|-----------------|
| 1 | DSC         | 92.3            |
| 2 | Sensitivity | 96.5            |
| 3 | PPV         | 90.4            |
| 4 | IoU (Train) | 92.94           |
| 5 | IoU (Test)  | 92.93           |

After updating the model parameters by 10 training iterations, all of the selected Alzheimer data were segmented in a 3D space by the proposed CNN algorithm. The segmented results were compared and evaluated with the manual segmentation by measuring the similarity metrics as shown in table 2. To better understanding the segmentation accuracy, the border of the segmented hippocampus was mapped to the original image in the axial, coronal and sagittal views as shown in figure 3.

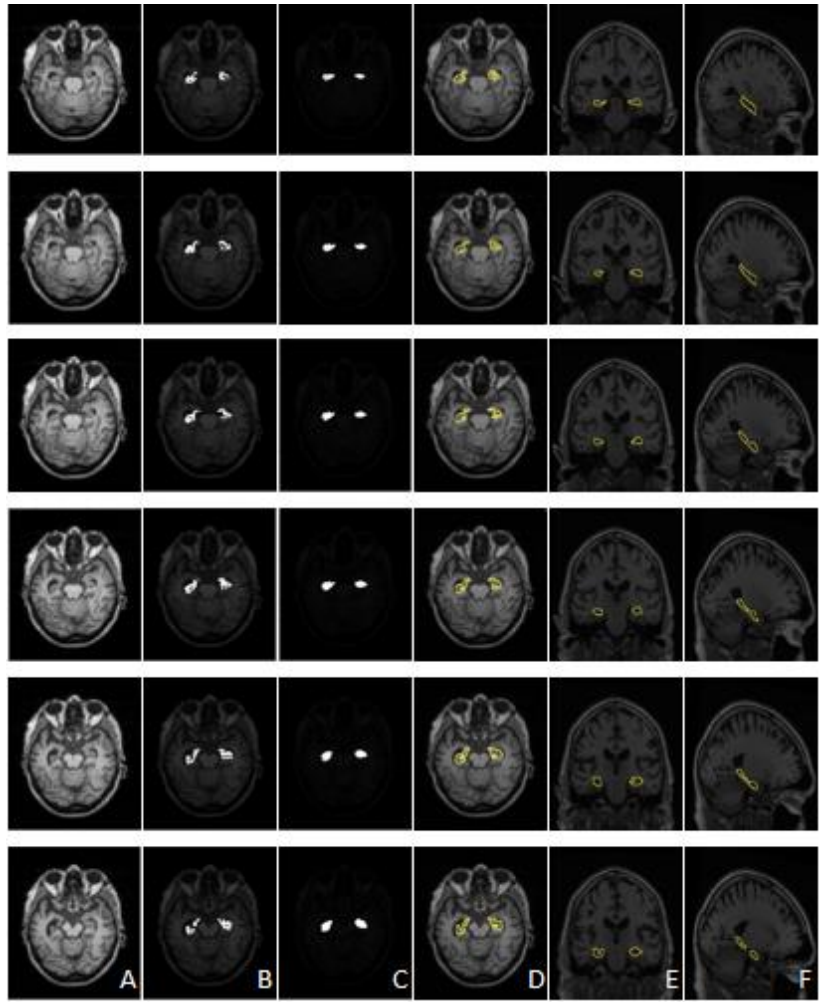

**Figure 3:** Segmentation results of the proposed method. A: Original images of the hippocampus area, B: The hippocampus mask added to the original images, C: Deep learning segmentation of the hippocampus area, D: The hippocampus border in the transvers plane obtained from CNN, E: The hippocampus border in the coronal plane obtained from CNN, F: The hippocampus border in the sagittal plane obtained from CNN.

**Discussion**

In this study, we proposed a CNN method to segment the hippocampus on the MR images of the Alzheimer's patients which publicly available from ANDI. We found that the segmentation results of the proposed method correlated with the manual segmentation which is considered as a gold standard. The results supported that our earlier hypothesis of CNN algorithms has superiority compared to other segmentation methods. The segmentation results were evaluated by computing the five similarity metrics and then, the performance of the proposed method was compared with several studies which focused on the hippocampus segmentation.

Table 3 shows comparison of the proposed method with some most used methods. The Dice metrics and computation times of the methods are seen in the table. Our proposed method showed better performance than others, but due to the complexity, the method had a little more computation time than some of the methods. However, the computation time of the model was shorter than that of the commonly used methods such as FreeSurfer and FIRST FSL.

The results are considered acceptable; indicating the success of our deep convolutional network training and the high accuracy of the hippocampal detection by neural networks which are called convoluted neural networks.

Table 3: Comparison of the proposed method with some most used methods.

| Study | Method | Dice Metric | Computation Time |
|---|---|---|---|
| Fischl et al. [26] | FreeSurfer (V 6.0) | 0.6980 | |
| Patenaude et al. [27] | FIRSTFSL (V 6.0) | 0.8044 | 10 min |
| Giraud et al. [28] | Optimized PAtchMatch Label Fusion (OPAL) | 0.898 | 0.92 s |
| Chen et al. [20] | U-Seg-Net + CLSTM | 0.8929 | - |
| Ataloglou et al. [29] | Deep-CNN | 0.9010　0.8835 | 14.8 s　21.8 s |
| This study | Deep CNN (U-Net) | 0.9230 | 323.4 s |

Our approach has some limitations such as: 1) if we had more data, the training and subsequent test results would be more accurate; 2) the number of convolution layers was high which increased the cost and complexity of the method; 3) In the current study some of data gives accurate results than others; the worst data demonstrates some of the challenging features for segmentation including the hippocampus shrinkage, mal-rotation and increased CSF around the hippocampus.

**Conclusion**

In this study, we proposed a method for segmenting the hippocampal area of the brain on MRI T1w images of the Alzheimer's patients. The core of our work was the use of Convolutional Neural Network (CNN), which has shown remarkable progress in machine learning. The neural network was trained using a supervised learning system. The training data was included MR images of the 100 datasets, described in more detail in previous sections, which were done manually by the hippocampal segmentation. Lastly, the network segmentation was evaluated as a test compared to the manual segmentation containing patient data. The proposed method had a better performance than others with acceptable computation time of 323.4 seconds. Therefore, the proposed method has superiority confidence related to others and can be used for the hippocampus segmentation.

**References**

1. Goedert M, Spillantini MG. A century of Alzheimer's disease. *Science*. 2006;**314**(5800):777-81. doi: 10.1126/science.1132814.
2. Kumar A, Singh A. A review on Alzheimer's disease pathophysiology and its management: an update. *Pharmacol Rep*. 2015;**67**(2):195-203. doi: 10.1016/j.pharep.2014.09.004.
3. Mu Y, Gage FH. Adult hippocampal neurogenesis and its role in Alzheimer's disease. Mol. *Neurodegener*. 2011;**6**(1):85. doi: 10.1186/1750-1326-6-85.
4. Braak H, Braak EV. Staging of Alzheimer's disease-related neurofibrillary changes. *Neurobiol Aging*. 1995;**16**(3):271-8. doi: 10.1016/0197-4580(95)00021-6.
5. Worker A, Dima D, Combes A, Crum WR, Streffer J, Einstein S, et al. Test–retest reliability and longitudinal analysis of automated hippocampal subregion volumes in healthy ageing and A lzheimer's disease populations. *Hum Brain Mapp*. 2018;**39**(4):1743-54. doi: 10.1002/hbm.23948.
6. Cao L, Li L, Zheng J, Fan X, Yin F, Shen H, et al. Multi-task neural networks for joint hippocampus segmentation and clinical score regression. *Multimed Tools Appl*. 2018;**77**(22):29669-86. doi: 10.1007/s11042-017-5581-1.

7. Dubois B, Feldman HH, Jacova C, DeKosky ST, Barberger-Gateau P, Cummings J, et al. Research criteria for the diagnosis of Alzheimer's disease: revising the NINCDS–ADRDA criteria. *Lancet Neurol*. 2007;**6**(8):734-46. doi: 10.1016/S1474-4422(07)70178-3.
8. Pennanen C, Kivipelto M, Tuomainen S, Hartikainen P, Hänninen T, Laakso MP, et al. Hippocampus and entorhinal cortex in mild cognitive impairment and early AD. *Neurobiol Aging*. 2004;**25**(3):303-10. doi: 10.1016/S0197-4580(03)00084-8.
9. Dill V, Franco AR, Pinho MS. Automated methods for hippocampus segmentation: the evolution and a review of the state of the art. *Neuroinformatics*. 2015;**13**(2):133-50. doi: 10.1007/s12021-014-9243-4.
10. Zhang J, Liu M, Shen D. Detecting anatomical landmarks from limited medical imaging data using two-stage task-oriented deep neural networks. *IEEE Trans Image Process*. 2017;**26**(10):4753-64. doi: 10.1109/TIP.2017.2721106.
11. Zhang J, Gao Y, Wang L, Tang Z, Xia JJ, Shen D. Automatic craniomaxillofacial landmark digitization via segmentation-guided partially-joint regression forest model and multiscale statistical features. *IEEE Trans Biomed Eng*. 2015;**63**(9):1820-9. doi: 10.1109/TBME.2015.2503421.
12. Hammers A, Allom R, Koepp MJ, Free SL, Myers R, Lemieux L, et al. Three-dimensional maximum probability atlas of the human brain, with particular reference to the temporal lobe. *Hum Brain Mapp*. 2003;**19**(4):224-47. doi: 10.1002/hbm.10123.
13. Litjens G, Kooi T, Bejnordi BE, Setio AA, Ciompi F, Ghafoorian M, et al. A survey on deep learning in medical image analysis. *Med Image Anal*. 2017;**42**:60-88. doi: 10.1016/j.media.2017.07.005.
14. Goodfellow I, Bengio Y, Courville A, Bengio Y. Deep learning. *Cambridge*, *MIT press*, 2016. doi: 10.4258/hir.2016.22.4.351.
15. Pan Y, Huang W, Lin Z, Zhu W, Zhou J, Wong J, et al. Brain tumor grading based on neural networks and convolutional neural networks. 37th Annual International Conference of the IEEE Engineering in Medicine and Biology Society (EMBC). 2015;699-702. IEEE. doi: 10.1109/EMBC.2015.7318458.
16. Zhao L, Jia K. Multiscale CNNs for brain tumor segmentation and diagnosis. Comput *Math Methods Med*. 2016;**2016**. doi: 10.1155/2016/8356294.

17. Dou Q, Chen H, Yu L, Zhao L, Qin J, Wang D, et al. Automatic detection of cerebral microbleeds from MR images via 3D convolutional neural networks. *IEEE Trans Med Imaging*. 2016;**35**(5):1182-95. doi: 10.1109/TMI.2016.2528129.
18. Wang C, Elazab A, Wu J, Hu Q. Lung nodule classification using deep feature fusion in chest radiography. *Comput Med Imaging Graph*. 2017;**57**:10-8. doi: 10.1016/j.compmedimag.2016.11.004.
19. Guo Y, Wu G, Commander LA, Szary S, Jewells V, Lin W, et al. Segmenting hippocampus from infant brains by sparse patch matching with deep-learned features. International Conference on Medical Image Computing and Computer-Assisted Intervention. 2014; 308-15. Springer, Cham. doi: 10.1007/978-3-319-10470-6_39.
20. Chen Y, Shi B, Wang Z, Sun T, Smith CD, Liu J. Accurate and consistent hippocampus segmentation through convolutional LSTM and view ensemble. International Workshop on Machine Learning in Medical Imaging, 2017;88-96. Springer, Cham. doi: 10.1007/978-3-319-67389-9_11.
21. Thyreau B, Sato K, Fukuda H, Taki Y. Segmentation of the hippocampus by transferring algorithmic knowledge for large cohort processing. *Med Image Anal*. 2018;**43**:24-28. doi: 10.1016/j.media.2017.11.004.
22. Boccardi M, Bocchetta M, Morency FC, Collins DL, Nishikawa M, Ganzola R, et al. Training labels for hippocampal segmentation based on the EADC-ADNI harmonized hippocampal protocol. *Alzheimers Dement*.2015;**11**(2):175-83. doi: 10.1016/j.jalz.2014.12.002.
23. Adaszewski S, Dukart J, Kherif F, Frackowiak R, Draganski B, Alzheimer's Disease Neuroimaging Initiative. How early can we predict Alzheimer's disease using computational anatomy? *Neurobiol Aging*. 2013;**34**(12):2815-26. doi: 10.1016/j.neurobiolaging.2013.06.015.
24. Taha AA, Hanbury A. Metrics for evaluating 3D medical image segmentation: analysis, selection, and tool. *BMC Med Imaging*. 2015;**15**(1):29. doi: 10.1186/s12880-015-0068-x.
25. Rahman MA, Wang Y. Optimizing intersection-over-union in deep neural networks for image segmentation. International Symposium on Visual Computing. 2016;**234**-44. Springer, Cham. doi:10.1007/978-3-319-50835-1_22.

26. Fischl B, Salat DH, Busa E, Albert M, Dieterich M, Haselgrove C, et al. Whole brain segmentation: automated labeling of neuroanatomical structures in the human brain. *Neuron*. 2002;**33**(3):341-55. doi: 10.1016/S0896-6273(02)00569-X
27. Patenaude B, Smith SM, Kennedy DN, Jenkinson M. A Bayesian model of shape and appearance for subcortical brain segmentation. *Neuroimage*. 2011;**56**(3):907-22. doi: 10.1016/j.neuroimage.2011.02.046.
28. Giraud R, Ta VT, Papadakis N, Manjón JV, Collins DL, Coupé P, Alzheimer's Disease Neuroimaging Initiative. An optimized patchmatch for multi-scale and multi-feature label fusion. *Neuroimage*. 2016;**124**:770-82. doi: 10.1016/j.neuroimage.2015.07.076.
29. Ataloglou D, Dimou A, Zarpalas D, Daras P. Fast and precise hippocampus segmentation through deep convolutional neural network ensembles and transfer learning. *Neuroinformatics*. 2019;**17**(4):563-82. doi: 10.1007/s12021-019-09417-y.